\pdfoutput=1
\documentclass[pdftex,twocolumn,10pt,letterpaper]{extarticle}

\usepackage[%
colorlinks=true
]{hyperref}

\usepackage{booktabs} % For formal tables

\usepackage{chngpage}

\usepackage{ifthen}
\usepackage{datetime}
\newcommand{\finalSubmission}{yes}

\ifthenelse{\equal{\finalSubmission}{yes}}{
        
    \newcommand{\showComments}{no}
}{
    \newcommand{\showComments}{yes}
}

\usepackage[normalem]{ulem}
\usepackage{soul}
\usepackage{xcolor}
\usepackage{url}
\usepackage{bm}
\usepackage{microtype}
\usepackage{placeins}
\usepackage{comment}
\usepackage{algorithm,algpseudocode}
\usepackage{amsmath}
\usepackage{mathabx}
\usepackage{flushend}

\usepackage{graphicx}
\graphicspath{{figures/}}

\newcommand{\note}[2]{
	\ifthenelse{\equal{\showComments}{yes}}{\textcolor{#1}{#2}}{}
}
\newcommand{\fixme}[1]{
	\ifthenelse{\equal{\showComments}{yes}}{\textcolor{red}{#1}}{#1}
}

\usepackage[font=bf]{caption}
\usepackage{subcaption}
\usepackage{xcolor}

\definecolor{worange}{RGB}{245, 128, 37}
\definecolor{darkgreen}{RGB}{0,100,0}
\definecolor{forestgreen}{RGB}{34,139,34}
\newcommand{\kl}[1]{\note{red}{[KL: #1]}}

\usepackage[numbers,sort]{natbib}

\usepackage{xspace}
\newcommand{\graphi}{\textit{Graphi}\xspace}
\newcommand{\nvidia}{Nvidia\xspace}

\author{{Linpeng Tang$^\coasterisk$, Yida Wang$^{\dagger}$,
		 Theodore L. Willke$^{\dagger}$, Kai Li$^{\coasterisk}$}\\
	{\small $^\coasterisk$Princeton University, $^\dagger$Intel Corporation}}
\title{Scheduling Computation Graphs of Deep Learning Models on Manycore CPUs}

\date{}

\begin{document}
%avoid overflow lines
\sloppy

\maketitle
	
%\date{}	
%\thispagestyle{empty}
%\showCurrentTime

\begin{abstract}
\vspace{0.2em}
%Deep learning frameworks typically run on the many-core processors for parallelism. 
For a deep learning model, efficient execution of its computation graph is key to achieving high performance. Previous work has focused on improving the performance for individual nodes of the computation graph, while ignoring the parallelization of the graph as a whole. However, we observe that running multiple operations simultaneously without interference is critical to efficiently perform parallelizable  small operations. The attempt of executing the computation graph in parallel in deep learning frameworks usually involves much resource contention among concurrent operations, leading to inferior performance on manycore CPUs. To address these issues, in this paper, we propose \graphi, a generic and high-performance execution engine to efficiently execute a computation graph in parallel on manycore CPUs. Specifically, \graphi minimizes the interference on both software/hardware resources, discovers the best parallel setting with a profiler, and further optimizes graph execution with the critical-path first scheduling.
%We have designed and implemented \graphi using intelligent scheduling algorithms with a profiler and a thread management system.
Our experiments show that the parallel execution consistently outperforms the sequential one. The training times on four different neural networks with \graphi are 2.1$\times$ to 9.5$\times$ faster than those with TensorFlow on a 68-core Intel Xeon Phi processor.
%\lt{note: we have comparable performance on small problems, but lower performance on large problems}

\end{abstract}

\section{Introduction}
\label{sec:intro}

\begin{comment}
As the contemporary computation moves towards more and more parallel processing, many-core processors featuring a large number of simple and independent cores play an increasingly important role in the computation. Typical many-core processors include graph processing units (GPUs) and Intel Xeon Phi processors (many-core CPUs).
\end{comment}

Manycore processor architectures utilize many relatively low performance cores to achieve high overall performance~\cite{Seiler:2008:LMX:1399504.1360617,vajda2011programming}. The architecture is particularly well-suited to high-performance computing (HPC) applications with lots of data parallelism, due to its large number of computing cores and wide vector processing units. One such application is deep learning~\cite{lecun2015deep}, whose models can be expressed as computation graphs with nodes representing the operations and edges representing the dependencies between nodes~\cite{bengio2009learning} (more details in Section~\ref{sec:background}).

The efficiency of processing computation graphs on contemporary computing devices, especially graphic processing units (GPUs), has been extensively studied in the literature\cite{krizhevsky2012imagenet,Sermanet_overfeat:integrated,chetlur2014cudnn,szegedy2015going}. Among these, many have focused on building efficient primitives to speed up single operations on one processor~\cite{wang2014intel,heinecke2016libxsmm} or optimizing distributed execution across multiple processors with a server~\cite{chen2015mxnet, hadjis2016omnivore} and across a cluster~\cite{parameterserver, xing2015petuum}. So far, little effort has been put into the scheduling of computation graphs on manycore processors.
Some previous methods use one executor to run a computation graph operation-by-operation sequentially on GPUs~\cite{bergstra2010theano, jia2014caffe}; others use a naive way to allow multiple executors to run simultaneously~\cite{abadi2016tensorflow,chen2015mxnet}, which introduces contention between threads sharing computing and memory resources. These approaches result in the substantial under-utilization of CPUs, the most popularly available computing resource, on the deep learning workloads.

In this paper, we study how to efficiently execute computation graphs of deep learning models on manycore CPUs.  Our experimental hardware platform is the Intel Xeon Phi processor, based on the Intel Many Integrated Core architecture (MIC) \cite{Seiler:2008:LMX:1399504.1360617}. We show that sequential execution normally cannot exhaust the available resource of this processor and that naive parallel execution typically achieves poor performance mainly due to sub-optimal thread scheduling and thread interference. Based on these observations, we propose \graphi, a generic high-performance execution engine for computation graphs on manycore CPUs. Our key idea is to profile a given computation graph, allocate resources to different agents (scheduler and executors) of the execution engine using the profiling results, and schedule operations intelligently with minimal interference. We compare running our execution engine on the manycore CPU with TensorFlow~\cite{abadi2016tensorflow} on the same hardware. 

To the best of our knowledge, \graphi is the first high-performance parallel execution engine with intelligent scheduling strategies for deep learning computation graphs on manycore CPUs. Although many of the techniques in \graphi are not new, our unique contribution is to identify issues of current deep learning frameworks on manycore CPUs, adapt proper techniques to a complex deep learning system, and make them work in synergy to greatly boost the overall performance. Moreover, we believe the concepts captured by \graphi can be incorporated into mainstream frameworks. Specifically, this paper makes the following contributions: \begin{enumerate}
\item We demonstrate that by choosing proper parallelism scheme and using optimized scheduling, \graphi outperforms TensorFlow on the Intel Xeon Phi processor by 2.1$\times$ to 9.5$\times$ on 4 popular deep learning networks.
\item We demonstrate that operations typically used in deep learning models (e.g. matrix multiplication and element-wise operation) saturate at 8 or 16 cores on the Intel Xeon Phi processor, and parallelizing multiple operations without unnecessary thread interference is preferable. We show that parallel execution outperforms sequential by up to 3.4$\times$.
\item We demonstrate that using a centralized scheduler to impose intelligent scheduling and eliminate software resource contention between autonomous executors further boosts overall performance of the execution engine by up to 19\%.
\end{enumerate}

 The rest of the paper is organized as follows: Section~\ref{sec:background} provides background on computation graphs and manycore CPUs, Section~\ref{sec:motivation} delves deeper into the motivation for our work and the challenges we have via microbenchmarking of the manycore CPU. We present the overall design of \graphi in Section~\ref{sec:design} and its implementation in Section~\ref{sec:impl}. Section~\ref{sec:other} discusses other optimization we considered during the system design. The evaluation is in Section~\ref{sec:eval}, followed by the discussion of related work in Section~\ref{sec:related}. Section~\ref{sec:concl} summarizes the paper and proposes the future work.

\section{Background}
\label{sec:background}

This section discusses the background of the two main aspects of the paper: computation graphs and manycore CPUs.

\noindent\textbf{Computation graph}
The computation graph is a common way to specify computation tasks and their dependencies for execution~\cite{karp1966properties}. This abstraction has found wide usage in dataflow computation~\cite{Dennis:1974:PAB:642089.642111,dennis1980data,tera,hep} and streaming data processing~\cite{carbone2015apache,toshniwal2014storm}. Recently, deep learning frameworks such as TensorFlow~\cite{abadi2016tensorflow}, MXNet~\cite{chen2015mxnet}, neon~\cite{neon}, Theano~\cite{bergstra2010theano} and Caffe~\cite{jia2014caffe} have used computation graphs to represent the required computation of deep learning models after compilation. A computation graph is a directed acyclic graph (DAG) with each node representing an operation which could be a matrix multiplication, a convolution or an element-wise operation, etc. A directed edge pointing from node A to B represents that operation B is dependent on operation A, i.e. the output of operation A serves as (part of) the input of operation B.

The training and inference of a deep learning model is essentially the execution of the corresponding computation graph. Training requires a larger computation graph which consists of both forward operations for computing the loss and backward operations for computing the gradients.  A complete execution on the graph corresponds to one training iteration of a batch. The computation graph for inference is smaller since it only contains the forward operations. One complete execution of the graph typically results in the inference of a group of instances.

The execution engine of a computation graph uses an important abstraction {\it executor} to lead a team of threads to run an operation at a time.  The team of threads allows the executor to use thread-level parallelism to execute the operation efficiently.  The size of the team can be configured or adjusted by the execution engine.

The conventional way of interpreting a computation graph is to execute operations in sequence according to a topological order of the graph. That is, starting from an operation with no dependencies (i.e. no other operations pointing to it), the execution engine picks one executable operation at a time to run. An operation is executable only after all operations pointing to it have finished (if any). 

The sequential execution approach requires only one executor, and improves performance by exploiting the parallelism within each operation to utilize the available SIMD and multi-thread parallelism of a CPU or GPU. This method works well when the computation graphs have large operations and simple structures (e.g., AlexNet~\cite{krizhevsky2012imagenet}).

A more advanced way is to execute operations with multiple executors in parallel when necessary, which is under explored and not optimized in popular frameworks. It holds the promise for improving the overall performance of computation graph execution on many complex networks.

\begin{figure}[tbp]
     \begin{center}
         \includegraphics[width=0.9\columnwidth]{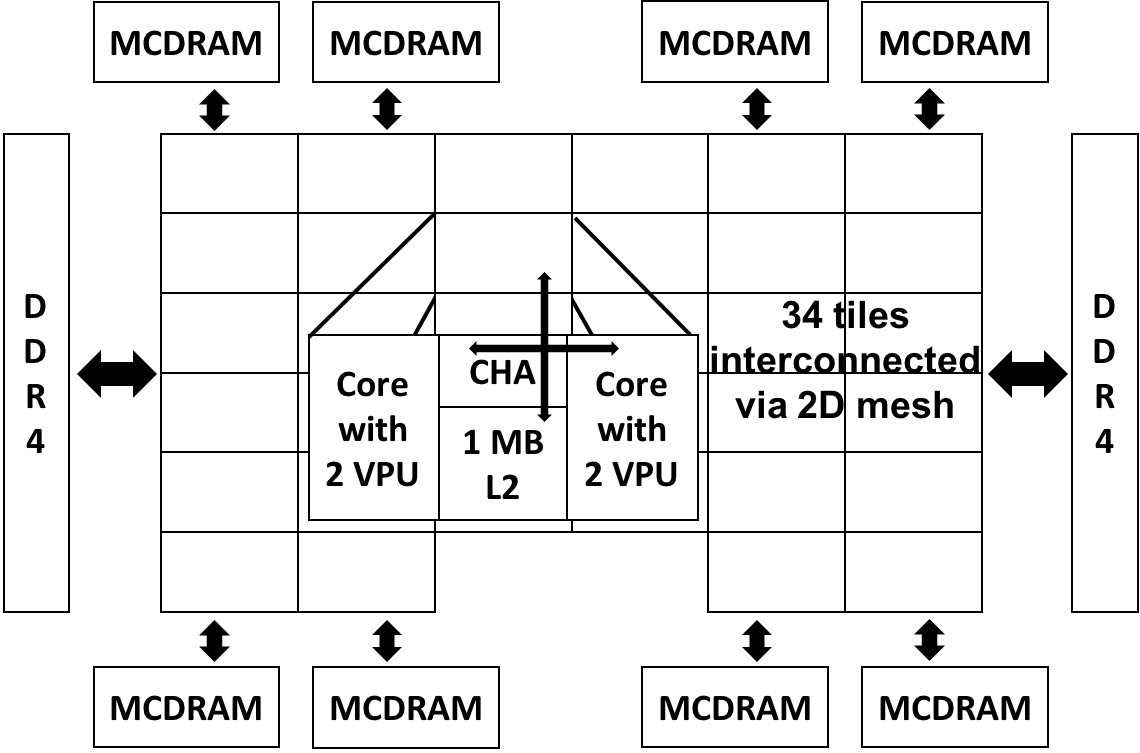}
         \caption{Architecture of Intel Xeon Phi processor 7250.}
         \label{fig:knl}
    \end{center}
    \vspace{-1em}
\end{figure}

%\kl{the following paragraphs provide a lot of details of Intel Xeon Phi.  The paper provides the detailed info again in evaluation.}\lt{We removed the details in evaluation.}

\noindent\textbf{Manycore CPU} This paper studies how to efficiently execute a computation graph on a manycore CPU.  Our experimental hardware is a 68-core Intel Xeon Phi processor 7250 (code named Knights Landing or KNL, referred to hereinafter as the manycore CPU) based on the Intel Many Integrated Core (MIC) architecture. It allows the use of parallel programming toolkits such as OpenMP in the same way as programming on a typical multicore x86 processor. The processor runs at a clock frequency of 1.40~GHz and supports up to 4 hardware hyper-threads per core.  Our experiments used one thread per core to eliminate interference among hardware threads running on the same core while achieving good performance.

Each core has $32$~KB L1 data cache, and $32$~KB L1 instruction cache. Every two cores are organized as a \emph{tile} with $1$~MB shared unified L2 cache. All tiles are interconnected as a 2D mesh. Cache coherence is maintained via a distributed  directory provided. The mesh supports three modes of tile clustering (all-to-all, quadrant, and sub-NUMA clustering) to provide different
levels of memory address affinity for better overall performance in different use cases. These cluster modes aim to lower latencies and improve bandwidth by reducing the distance of data traversals within the chip~\cite{sodani2016knights}. 

In this paper, the manycore CPU is configured in the \emph{quadrant} mode, which offers symmetric memory access. For a more sophisticated system design, the manycore CPU may use the \emph{sub-NUMA clustering} mode for better performance. The manycore CPU is equipped with a 16~GB multi-channel DRAM (MCDRAM) with bandwidth greater than 400~GB/s, as well as 96~GB DDR4 memory. Figure~\ref{fig:knl} depicts the architecture. Understanding of this architecture helped to shape the design of our execution engine in Section~\ref{sec:design}.

\begin{figure*}[htbp]
	\centering
	\begin{subfigure}{0.48\textwidth}
		\centering
		\includegraphics[width=\textwidth]{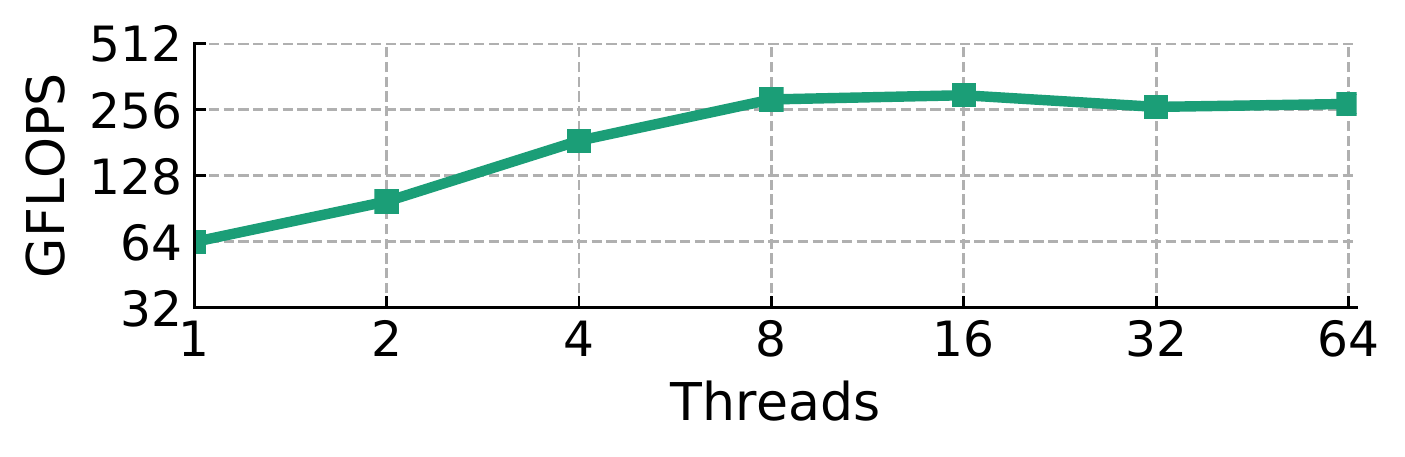}
		\caption{GEMM}
		\label{fig:gemm-scalability}
	\end{subfigure}
	~
	\begin{subfigure}{0.48\textwidth}
		\centering
		\includegraphics[width=\textwidth]{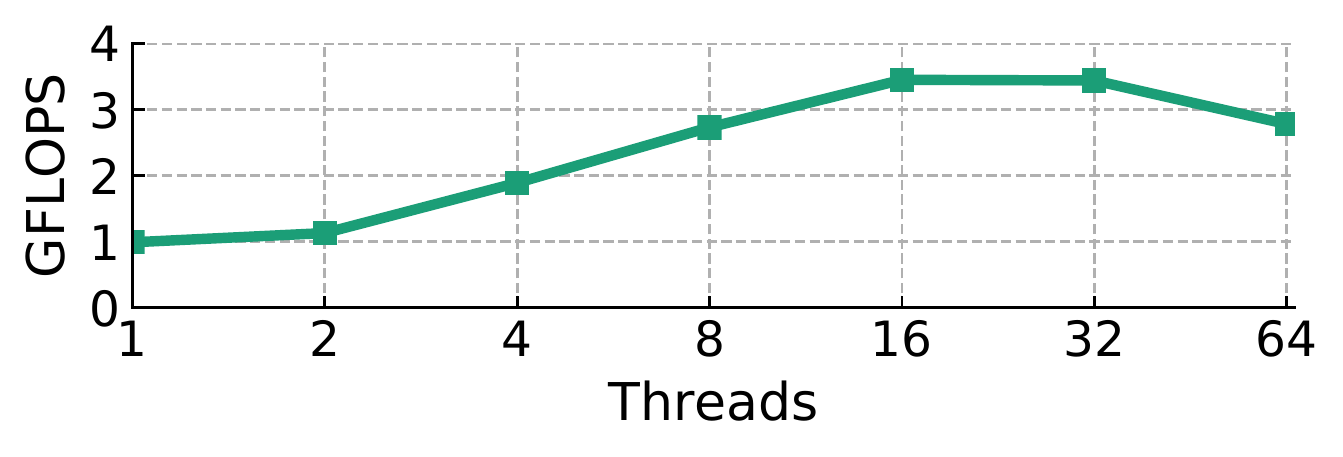}
		\caption{Element-wise Multiplication}
		\label{fig:elementwise-scalability}
	\end{subfigure}
	\caption{Scalability of GEMM/element-wise multiplication operations in a typical LSTM on Intel Xeon Phi processor 7250.}
	%\vspace{-1em}
	\label{fig:scalability}
\end{figure*}

\section{Motivations}
\label{sec:motivation}
%\lt{We need to explain better why single operations are slow. One cause is that the many threads running in a single thread can lead to stragglers. We can think about how to quantify these effects as well.}
%\lt{Generally, we should talk how our design alleviates/solves the problem of manycore: straggling/overhead solved with parallel executors, and weak cores solved by avoiding interference and scheduler design. Small cache size is yet to be solved :)}

This section first discusses the challenges of executing computation graphs on a manycore CPU efficiently, then performs microbenchmarking on the manycore CPU to further investigate the issues, which motivates the design of \graphi.

\subsection{Challenges}
\label{sec:motivation:challenge}
The conventional way of executing computation graphs in sequence does not work well on the manycore CPUs for networks with smaller operations and complex linking structures (e.g. long short-term memory (LSTM)~\cite{hochreiter1997long}, PathNet~\cite{fernando2017pathnet}). These networks have operations that are too small to fully take advantage of the compute power of all cores because of the thread management overhead, which is getting worse as the number of cores increases. In order to obtain better resource utilization, multiple operations should be run in parallel. % We will verify this later in Section~\ref{sec:motivation:study}.

Fortunately, independent operations in a computation graph are always parallelizable. Modern frameworks like TensorFlow and MXNet provide parallel execution engines that can execute more than one operation at the same time. Nevertheless, these frameworks have not been carefully optimized for manycore CPUs, and the challenges described below can hinder their execution efficiency on the hardware.

The first challenge is how to optimally schedule the operations of a computation graph on a number of operation executors. Scheduling $M$ operations expressed as a DAG to run on $N$ executors to minimize the \emph{makespan}, the total time from execution starts until every operation finishes, is a well-known NP-hard problem by reduction from the 3-partition problem~\cite{leung2004handbook}. 

The approach TensorFlow and MXNet have taken is to maintain a centralized queue of the executable operations without dependencies, and allows an arbitrary executor to execute any operation that is at the head of the queue. Once the dependencies of an operation are all executed, it will be placed onto the queue. The scheduling continues until all operations are executed. This naive scheduling strategy is simple but may not perform well on the manycore CPU.

Another challenge is how to eliminate the interference among executors, which is commonly seen in the modern deep learning frameworks, leading to performance reduction of the parallel execution engine and even decreases the overall performance compared with sequential execution.

\begin{comment}
For example, multiple executors running on a GPU as different compute streams share a global L2 cache but all assume they exclusively have the entire resource. This may cause contention and performance loss. Therefore, by now some popular deep learning framework like TensorFlow does not implement parallel execution engine for GPUs. 

\kl{This paper talks about CPU, not GPU.  Perhaps no need for this example.}
\end{comment}

For instance, TensorFlow and Caffe2 (an updated version of Caffe)~\cite{caffe2} use Eigen Library~\cite{eigen} as well as OpenMP for different operations, each with their own thread pool, which results in more software threads than available physical cores. This over-subscription causes either unnecessary resource contention or expensive thread context switching on the manycore CPU. Moreover, those frameworks do not explicitly specify on which cores the threads should run, making it likely for execution threads to compete for the same physical cores.  Such contention can cause threads to straggle, and further hampers overall performance.

Contention may happen over software resources as well. One example is the centralized queue of the executable operations. When the number of executors is large and the execution time of an operation is small, the overhead of global queue polling contention becomes significant. In general, contention on the manycore CPU could be very severe due to its large number of cores.

\begin{figure*}[htbp]
	\centering
	\begin{subfigure}{0.48\textwidth}
		\centering
		\includegraphics[width=\textwidth]{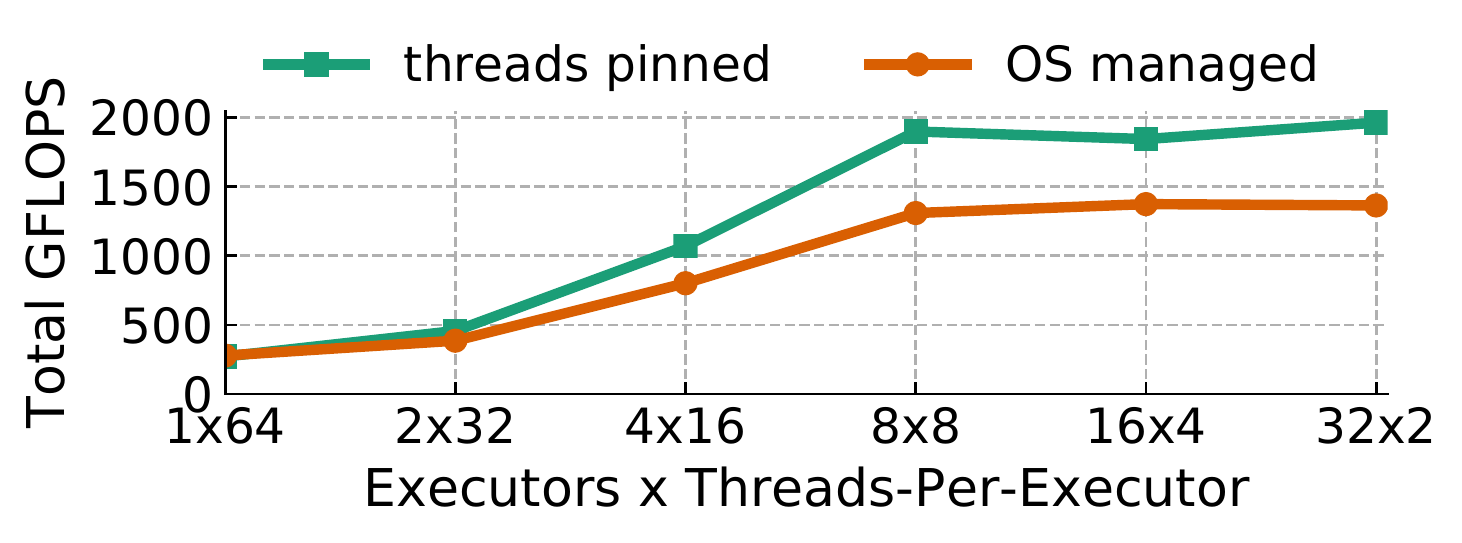}
		\caption{GEMM}
		\label{fig:gemm-interference}
	\end{subfigure}
	\vspace{-0.5em}
	~
	\begin{subfigure}{0.48\textwidth}
		\centering
		\includegraphics[width=\textwidth]{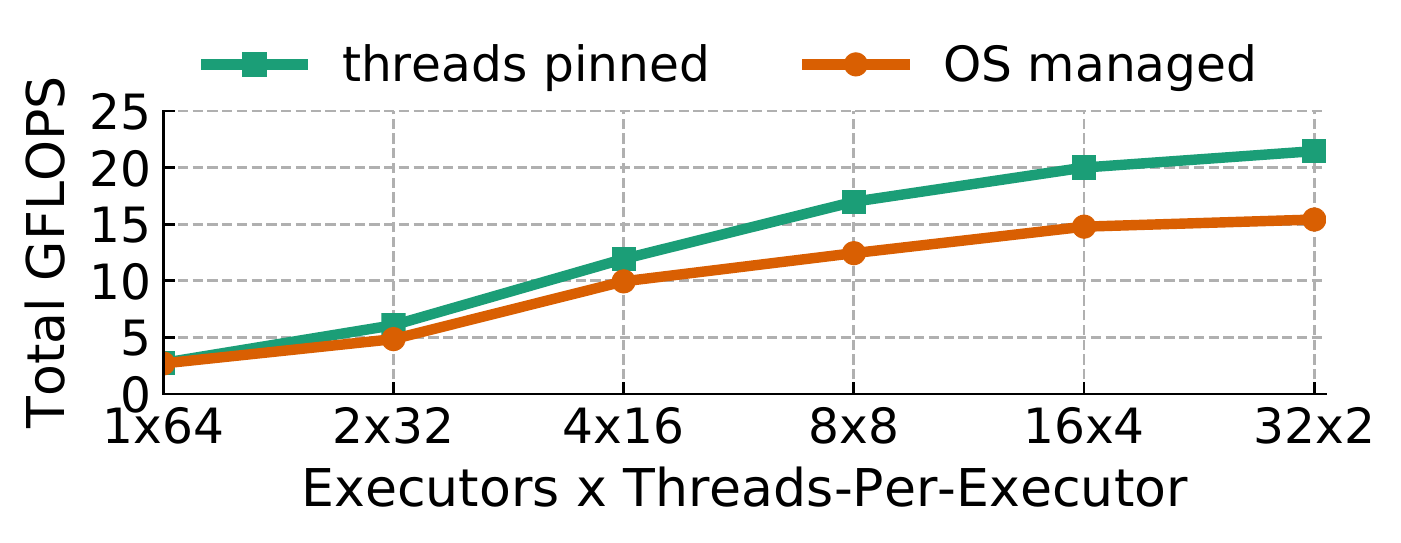}
		\caption{Element-wise Multiplication}
		\label{fig:elementwise-interference}
	\end{subfigure}
	\caption{Performance of parallel operations with pinned vs. OS managed threads on Intel Xeon Phi processor 7250. There are multiple executors, each running GEMM/element-wise multiplication operations with a team of threads.}
	%\caption{Performance of parallel GEMM/element-wise multiplication operations with pinned vs. OS managed threads on Intel Xeon Phi processor 7250.}
	%\vspace{-1em}
	\label{fig:interference}
\end{figure*}

\subsection{Microbenchmark Performance}
\label{sec:motivation:study}

We designed two microbenchmarks to validate the following two performance characteristics of the manycore CPU: 1) the scalability of small operations degrades at some point and 2) running multiple small operations in parallel without interference is beneficial. Although these concepts are well-studied in general, we do not see them embodied for typical deep learning operations on the manycore CPU. Therefore, we find it critical to validate them before designing a scheduling system for deep learning workloads on the manycore CPU. 

The first microbenchmark was used to assess the scalability of a manycore CPU for small operations. This benchmark includes two commonly used operations in the computation graph of LSTM: matrix multiplication (GEMM) of size $[64, 512] \times [512, 512]$ implemented via Intel Math Kernel Library (MKL), and element-wise multiplication for $32~ 768$ element pairs multi-threaded via OpenMP. The specific sizes are chosen to represent the medium size of LSTM suggested in the standard TensorFlow benchmark. 

Figure~\ref{fig:scalability} shows that the performance of GEMM saturates when the number of threads is greater than 8, whereas the element-wise multiplication saturates when it is greater than 16. Therefore, dedicating all available computing resources of the manycore CPU to a single operation is not optimal and the parallel computing power is largely wasted.

The second microbenchmark was designed to show the effect of resource contention within the manycore CPU. This microbenchmark consists of multiple GEMM and element-wise multiplication instances.  The sizes of these two operations are the same as those in the first microbenchmark.  We ran the microbenchmark in two modes on the same manycore CPU: manually pinning different threads to different physical cores and leaving the thread assignment to OS.

Figure~\ref{fig:interference} shows that the overall FLOPS of operations with threads pinned is higher than OS managed by up to 45\%. This is because the OS is unaware of the layout of the physical cores of the manycore CPU, so it is likely that multiple threads ran on the same physical core, which can cause synchronization overheads and cache misses.  Since a modern manycore CPU has private caches within each tile, the scheduling of executors on physical cores should be architecturally aware to reduce such cache contention.  Pinning threads to cores properly removes such overheads.  

By comparing the peak FLOPS in Figure~\ref{fig:scalability} and Figure~\ref{fig:interference}, we can see that the overall performance of running multiple small operations together without interference is more than 6$\times$ faster than running one single small operation using all the available resources. Such results validate the value of running multiple small operations in parallel.

Based on our benchmark experiments, we argue that an efficient parallel execution engine should avoid contentions on the manycore CPU and should execute small operations of a computation graph in parallel.

\section{\graphi Design}
\label{sec:design}
We propose \graphi, a generic and high-performance execution engine to efficiently run computation graphs of deep learning models on the manycore CPU. \graphi has multiple kinds of agents, e.g. profiler, scheduler, and executor, whose functions will be described in detail later. We keep the following goals in mind while designing the system:
\begin{enumerate}
\item The system should be general purpose, and be able to execute different kinds of neural networks;
\item Given a computation graph, the system should be able to schedule and execute the operations in a way that minimizes the \emph{makespan}.%, i.e., the total time of finishing all operations; \lt{makespan already defined}
\item In presence of a fleet of multi-threaded executors and a centralized scheduler, the system should avoid interference across these agents.

\end{enumerate}

%\kl{Agent is not defined.  Perhaps the architecture diagram should show agents?}

\subsection{Overview}
\label{sec:design:overview}
\begin{figure}[htbp]
     \begin{center}
         \includegraphics[width=0.9\columnwidth]{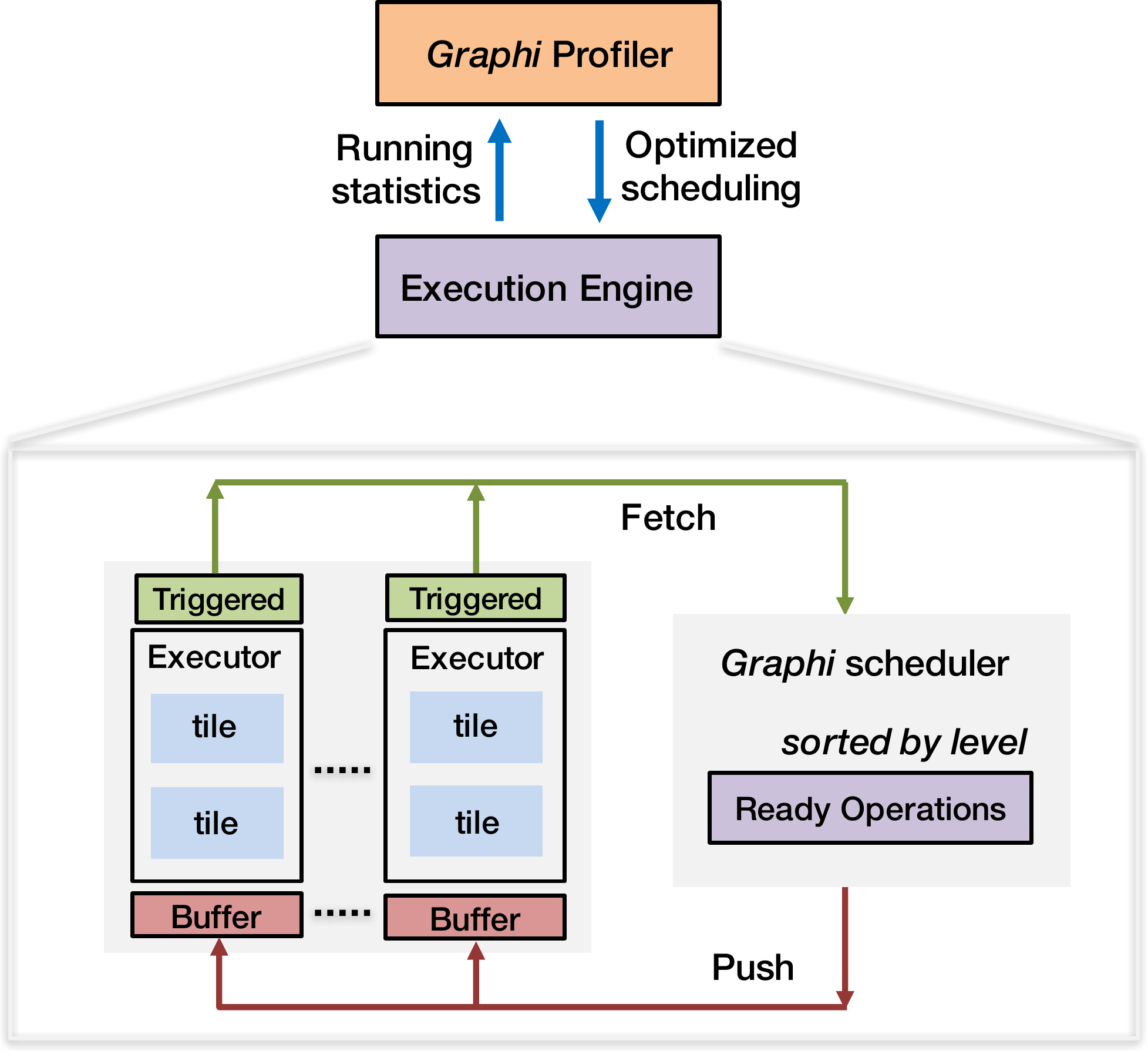}
         \caption{\graphi design overview.}
         \label{fig:graphi-overview}
    \end{center}
\end{figure}

Figure~\ref{fig:graphi-overview} shows the architecture of \graphi. \graphi has two kinds of inputs: a compiled computation graph of a deep learning model and the number of cores of a manycore CPU. We assume that the computation graph is static, meaning that the graph will not change during the entire computation.

%\kl{Perhaps the flow of this section should be top town.  First, we describe how Graphi works using the architecture diagram, components by components.  Then, we have a subsection for each component.}

In order to fulfill design goal (1), \graphi is designed to be agnostic to the underlying neural network, only seeing the computation graph as a DAG. For an arbitrary DAG, \graphi will first profile it for a better scheduling strategy during In the initial few runs. 

The \graphi profiler works with the execution engine in a feedback loop as shown in the upper part of Figure~\ref{fig:graphi-overview}: the execution engine runs the graph to generate statistics and informs the profiler, while the profiler uses these statistics to collect information and improve scheduling. The execution engine then uses this information to optimize subsequent runs, therefore fulfilling design goal (2). In general, \graphi works to avoid possible hardware and software resource contention following (3).

\subsection{Profiler}
\label{sec:design:profiler}

The \graphi profiler has two goals.  The first is to determine the basic configuration for execution.  Given the number of available cores, it comes up with different combinations of number of executors and threads per executor in order to find one with minimal execution makespan. For simplicity (see Section~\ref{sec:other} for more discussions), we make the executors symmetric, i.e. all executors own the same number of threads. In this way the profiler only needs to enumerate through a small number of configurations. For example, assume there are 64 threads, then we may have 1 executor with 64 threads, or 2 executors with 32 threads each, etc., up to 64 executors with 1 thread each. The \graphi execution engine will then use the selected combination for the subsequent optimizations.

The second goal is to estimate the running time of each operation of the computation graph with the selected combination. This adds modest overhead (i.e. execution time) to the system, but since the computation graph is static, this information can be treated as invariant and only needs to be collected for the first few iterations. It can then be used by the scheduler to further improve the scheduling process.

\subsection{Scheduler}
\label{sec:design:scheduler}

%\kl{I think this subsection needs improvements.  Many paragraphs try to justify the design by comparing with other approaches.  I think it is better to describe the data structures used and the algorithm first, and then argue why this design is better by telling the properties and by comparing with others.}

\graphi uses a centralized scheduler to coordinate the operations running on different executors, which simplifies the system processing pipeline by concentrating the scheduling decisions to a single agent. The lower part of Figure~\ref{fig:graphi-overview} illustrates the design of our scheduler, and Algorithm~\ref{alg:scheduler} summarizes its workflow.

The centralized scheduler oversees the execution status. It keeps polling for the newly triggered operations from the executors, as well as allocating ready-to-run operations to the executors. The scheduling is done once all operations are executed. Since the scheduler has knowledge of the system state, i.e. which operations are executable and their dependencies with the other operations, it can make strategy accordingly of sending which operation to run once an executor becomes available. One analogy of this design choice is the centralized software defined networks (SDN)~\cite{greenberg2005clean} compared with the traditional decentralized network protocols.

Although we know the graph structure and the estimated executing times of all operations by profiling the graph, an optimal offline scheduling solution is not feasible because there are unpredictable variations at run time. These variations can cause empty cycles spent waiting for operation dependencies. Therefore, we focus on an online scheduling solution.

We design the scheduler using an online algorithm~\cite{hu1961parallel} to prioritize the operations in the critical path of the DAG. Specifically, from the information of the computation graph structure and the estimated running time of each operation, we can derive a \emph{level} value for each operation, which is defined as the longest accumulated time from this operation to the end (sink point) of the computation graph. \graphi sorts the read-to-run operations according to their level values decreasingly and always schedules the operations with higher level value first. In other words, the operations in the critical path are prioritized for earlier execution so that they will not become the bottleneck. We call this \emph{critical-path first scheduling}.

\begin{algorithm}[htbp]
	\caption{\graphi Scheduler}
	\label{alg:scheduler}
	\begin{algorithmic}[1]
		\While {\textit{hasPendingOperations()}}
		\State Poll triggered operations from each executor
		\State Sort ready operations based on their \emph{level} values
		\While {\textit{hasReadyOperations()} and \textit{FindExecutor}($e$)}
		\State Get operation $p$ with the maximal \emph{level} value
		\State Put $p$ into executor $e$'s buffer
		\EndWhile
		\EndWhile
	\end{algorithmic}
\end{algorithm}

The \graphi scheduler design improves upon existing scheduling schemes in two ways. First, it eliminates potential software resource contention. In the parallel execution engines of TensorFlow and MXNet, there is only one queue that maintains the operations ready for execution (i.e., have no dependencies or all dependencies are satisfied). All executors independently poll the same queue for the next operations. This results in a heavy contention on the global queue, especially when the number of executors is large and the execution time of an operation is small. \graphi avoids this issue by having the scheduler push operations to executor-specific operation buffers. Since the buffers are disjoint, the interference between polling executors is eliminated.

The other advantage of the centralized scheduler is that it gives us flexibility to use different advanced scheduler polices. Current scheduling strategy is critical-path first, but the architecture allows us to easily implement other strategies. This is not possible in other state-of-the-art parallel execution engines, which lack a centralized scheduler. As discussed in Section~\ref{sec:motivation}, those execution engines schedule the operations in an arbitrary topological order, that is, whenever an executor is available, it randomly picks a ready operation to run. Since all executors work greedily, a global optimization strategy cannot be imposed.

\subsection{Executors}
\label{sec:design:executor}
In addition to the centralized scheduler, \graphi uses a fleet of executors harnessed by the scheduler. The number of executors, as well as the team size of threads per executor, are determined by the profiler. The executors are in charge of executing operations assigned to them by the scheduler. Its workflow is shown in Algorithm~\ref{alg:executor}. 

As discussed in Section~\ref{sec:design:scheduler}, the executors only need to poll their own operation buffers for operations to execute. In addition, to further reduce contention, each executor is also associated with its own triggered queue, where it outputs the triggered operations upon finishing one operation. These are then fetched by the scheduler for processing.

\begin{algorithm}[htbp]
	\caption{\graphi Executor}
	\label{alg:executor}
	\begin{algorithmic}[1]
		\While {true}
		\State /* Poll the buffer for new operation */
		\If {\textit{GetOperation}($p$)}
			\State Execute $p$ with the team of threads
			\State Trigger $p$'s depending operations
		\EndIf
		\EndWhile
	\end{algorithmic}
\end{algorithm}

In our design, we assign each executor exclusively to a number of tiles (see lower part of Figure~\ref{fig:graphi-overview}), each of which consists of two physical cores and an exclusive L2 cache as described in Section~\ref{sec:background}. As a result, executors do not share the compute units nor L2 cache, and consequently the hardware resource contention we discussed in Section~\ref{sec:motivation} is largely avoided.

%Since different executors runs on disjoint cores, they do not share compute unit, and consequently, L2 cache, the hardware resource contention we discussed in Section~\ref{sec:motivation} is largely avoided.
%\lt{Mention that we don't partition one tile among multiple executors. Ah, we also mention this later.}
\section{Implementation}
\label{sec:impl}

The implementation of \graphi leverages the computation graph toolkit (CGT)~\cite{cgt}. The main reasons for choosing CGT are its modular design of compilation and execution and its small code base. We use CGT's compilation component to compile a deep learning model into a computation graph. But, we add the profiler component, and completely redesign and reimplement the execution engine. These are the focuses of this paper, and the implementation can also be migrated to other deep learning frameworks. 

This section gives an overview of a typical deep learning framework like CGT, and then describes the implementation issues in \graphi.

\subsection{Overview of CGT}

%This section gives an overview of CGT.  Most of the designs covered herein can apply to other deep learning frameworks that use computation graphs. 

CGT consists of two main parts: 1) a \emph{compiler} to compile from a model into a computation graph and 2) an \emph{execution engine} to run the graph. The model definition is constructed in Python, represented as mathematical expressions relating to the inputs, the intermediate variables, and the outputs. The compiler translates high level expressions to a low level computation graph targeting a particular processor, in our case the manycore CPU. 

Each variable will be assigned a memory location, and optimizations during compilation allow multiple variables to share the same location as long as their lifespans do not overlap. Compute-intensive operations implemented in C++ are also compiled into shared libraries as callable routines. 

The execution engine, as well as the data structures of the computation graph, including variables, operations, and their dependencies, are also written in C++ for efficiency. The engine is discussed in more detail in the next susbsection.

% The computation graph structure is represented as a list of operations, each of which consists of the memory locations of the inputs and the outputs, as well as the pointer to the compiled function. All the related data structures as well as the execution engine is implemented in C++ for efficiency.\lt{remove this paragraph if necessary}

%The execution starts from the operations with no incoming edges, with the number of fulfilled incoming edges for every operation maintained during the process. An operation is triggered when all the incoming edges have been fulfilled, and can then be scheduled to run.
%\lt{Duplicate with other places.}

\subsection{Profiler and execution engine}

As described previously, \graphi has three components: the profiler, the centralized scheduler, and a fleet of executors. This subsection discusses their implementation details.

\textbf{Profiler.} After determining the best configuration for graph execution, the profiler records the information of each operation for several runs, which includes start and end time, the input/output data address and size, as well as the executor running it. The computed duration is averaged over multiple iterations to reduce variance, and then it is used in the critical-path first scheduling. The data addresses help analyze data locality (see Section~\ref{sec:other} for our attempts on improving locality). In addition, we use the profiling results to visualize the execution process , i.e. placing the operations to their running executors' timelines. This has been immensely helpful in analysis and debugging. Normally, the profiler only runs in the first few iterations, adding minimal overhead to a typical deep learning workload running for thousands of iterations.

%\lt{Shorten these two paragraphs after moving algorithm to design.}
\textbf{Scheduler.} We dedicate the client thread initiating the graph execution to run the scheduler, and it works in a busy loop. In each iteration, it first polls for the newly triggered operations. In order to do the critical-path first scheduling, it then maintains the operations in a \emph{max binary heap} ordered by their \emph{level} values, and these operations can be fired once an executor becomes available.

In order to efficiently check the available executors and make advanced scheduling decisions, the executor states are represented as a bit map, with 1 denoting the executor is idle, and 0 denoting busy. We use bit-scan intrinsics to find the number of trailing zeros in the bit map, which corresponds to the first executor now available to run. The scheduler then pushes the operation at the head of the heap to that executor's operation buffer. %For a many-core CPU, we reserve one core to the scheduler to avoid interference.\lt{remove rep}

\textbf{Executors.} Each executor polls for operation from its own operation buffer, executes the operation, and triggers this operation's dependencies. The operation buffer is implemented with a lock free ring buffer for high efficiency. This implementation is inspired by the per-thread run queue of MuQSS~\cite{muqss} scheduler, and enables us to buffer multiple operations to further reduce scheduling overheads and apply more sophisticated scheduling schemes.

In practice, we find that the load imbalance caused by a larger buffer size offset the benefit, so we buffer at most one operation in \graphi. Each executor spawns a team of OpenMP threads to run the operations. Our implementation chooses an even number of threads such that no two executors shared a tile (see Section~\ref{sec:background}), consequently avoiding L2 cache interference among executors. 

We use the primitives in several software packages including  LIBXSMM~\cite{heinecke2016libxsmm} for convolutions, Intel MKL for matrix multiplications, and OpenMP for loop of element-wise operations.  The engine uses OpenMP for thread management. As long as the size of the thread team of an executor does not change across different operations, the OpenMP library will always reuse the same team of threads, with the executor thread being the master. 

Before one executor launches, it creates an OpenMP parallel region for its team of threads, in which each thread in the team is pinned to a specific core. During the execution of subsequent operations, the thread will stay on the same core. We find this setting important for high performance because it eliminates resource contention as well as the overheads from thread migrations and context switches. 

\begin{comment}
Graphi execution engine can also use other thread packages as long as the threads can be shared by the operation building primitives and contention is eliminated, high performance is achievable.
\end{comment}

The executors discussed above are designed to run expensive operations such as convolutions, matrix multiplications, and large element-wise operations. In addition, there are also small operations like scalar addition in the computation graph. Both CGT and TensorFlow employ an optimization to directly run these small operations in the current thread/executor instead of pushing them to the ready-operation queue. We adopt the same idea in \graphi.

%\kl{Does Graphi uses the same approach as CGT and TensorFlow?}\lt{We do, added one sentence.}

It is also worth noting that bootstrapping the computation graph requires running some small operations. The state-of-the-art execution engines usually piggyback them to the framework's client thread that initiates the graph execution, but this hinders the scheduling process, which runs on the same thread. To solve this issue, \graphi maintains a light-weight single-threaded executor to take care of these operations. In order to avoid interference, one core is reserved exclusively for this executor as well.

%\kl{this paragraph is not clear}

\section{Optimization Considerations}
\label{sec:other}

%\lt{This subsection can be shortened.}

Since the design of \graphi allows us to implement many scheduling policies, we experimented several approaches. This section reports some of such optimization attempts and the insights we gained.

\paragraph{Different executor thread team sizes.}
A complex network such as LSTM consists of operations with varying sizes and different scalability. We performed a study to execute the computation graph with a sequential interpreter running on varying number of threads, and found that running time of operations scaled differently with the number of cores used. Based on this, we classified the operations into multiple classes (e.g. 3) according to how well they scale, and made the scheduler preferably assign an operation to an executor of corresponding thread team size.
%Based on this observation we classified the operations into multiple classes (e.g. 3) according to how well they scale. Then we configured \graphi to have executors of different thread team sizes. During scheduling, when there was an available executor, the scheduler first tried to assign to it an operation of corresponding scalability, before trying other operations. For example, for a larger executor employing more threads, the scheduler preferred assigning an operation that scales better and so on. 

This technique indeed reduced the total CPU time of all the threads. However, the makespan of the whole graph execution did not improve.  This was because different executor sizes could cause work straggling when some big operations are scheduled to run on the executors with a small team of threads. Therefore, the current \graphi uses symmetric executors with the same number of threads. Whether varying team sizes is useful on other models requires more investigations.

\paragraph{Dynamic number of executors.} We considered varying the number of executors dynamically during of the course of graph execution. For example, we tried to use different numbers of executors for forward and backward computations during a model training.   The rationale is that typically the number of parallel operations doubles during the backward pass.% because the gradient w.r.t. to the input of each layer as well as the parameter weights can be computed at the same time, whereas in the forward pass only the output of each layer needs to be computed.\lt{save space}

We found that two issues prevented this optimization method from being effective. First, there is a limitation with OpenMP such that thread reuse could not be guaranteed if the thread team size changed dynamically. Our experiments showed that the overhead of context switches between different threads on the manycore CPU is significant, at about 10-30 ms. This is aligned with the numbers provided in~\cite{barroso2017attack}. 
Second, as shown later in Figure~\ref{fig:parallelism}, after certain saturation points, increasing the parallelism only by a factor of $2$ reduces the overall running time only slightly.  The optimization to double the executors during the backward pass might not be worthwhile especially when the operations are large enough. 
%The tradeoff is a more complex design that requires more knowledge about the computation graph.

\paragraph{Data cache locality.} Pinning threads of an executor to specific cores gives control over the execution location of the threads. Combined with the knowledge from the centralized scheduler, we could naturally incorporate L2 cache locality in the \graphi execution engine. Note that when one operation finishes, it may trigger another using its result as the input data. In this case, we made the system remember the current executor as the \emph{preferred executor} for this triggered operation, which would then have the priority to run as the next operation on the preferred executor. Such cache affinity idea had been studied in the past~\cite{philbin1996thread,rogers2012cache}.

When analyzing the execution times of individual operations, we found that only element-wise operations improved by a modest margin, while matrix multiplications did not improve. Our hypothesis is that this is due to the blocking scheme of Intel MKL on the input/output matrices. As long as the threads of one executor does not totally reside in the shared L2 cache within a tile (consisting of two cores), data still have to traverse between different L2 caches, defeating the purpose of cache affinity scheduling. Further experiments where each thread team resided in one tile showed modest yet consistent improvement, confirming our hypothesis, but we did not pursue the idea because of this restrictive setting.

%This hypothesis was confirmed by our further experiment showing modest yet consistent performance improvements when each executor had a thread team of size 2 to occupy only one tile; they had no need to traverse data between L2 caches. However, this setting was rather restrictive, and we did not pursue it further. Nevertheless, such idea could be useful when the computation graph ran on multiple NUMA nodes (in fact, the many-core CPU has a sub-NUMA cluster mode), and we left this idea for future optimizations.

In contrary to locality, we find that writing through the results of an element-wise operation to memory with stream store\footnote{achieved through \texttt{\#pragma vector nontemporal}} slightly improves the overall performance.  Since it is likely that the results will not be reused by the same executor, there is no need to fetch the overwritten data into the cache to cause additional overheads. Therefore we adopt this technique in all our design.

\section{Evaluation}
\label{sec:eval}

We tested our implementation of \graphi with 4 representative deep learning models including LSTM, PhasedLSTM, PathNet and GoogleNet. Our evaluation seeks to answer the following key questions:

\begin{enumerate}
\item What's the overall performance of \graphi on the manycore CPU? How does it compare with the state-of-the-art deep learning framework like TensorFlow? (\S\ref{sec:eval:overall})
\item How much parallelism is needed for running different computation graphs on the manycore CPU, and what is the relationship between parallelism and performance? (\S\ref{sec:eval:parallel})
\item How much benefit does our centralized scheduler have on a hardware resource contention free baseline design? (\S\ref{sec:eval:scheduler})
\end{enumerate}

\subsection{Experiment Setup}
%\lt{Also mention we have implemented a simple OpenMP-based pooling layer.}

\begin{table*}[t]
	\begin{adjustwidth}{+0.1in}{+0.1in}
		\centering
		%	\begin{subtable}{0.5\textwidth}
		%		\centering
		%		\scalebox{1.0}{
		%		\begin{tabular}{l c c c}
		%			\midrule
		%			\textbf{Size} & \textbf{Sequence} & \textbf{Layers} & \textbf{Neurons} \\
		%			\hline\\[-1.75ex]
		%			Small & 20 & 4 & 128 \\
		%			Medium & 30 & 4 & 512 \\
		%			Large & 40 & 4 & 1024 \\
		%			\midrule
		%		\end{tabular}
		%	    }
		%		\caption{LSTM/PhasedLSTM}
		%	    \label{tbl:lstm-params}
		%	\end{subtable}%
		%	\begin{subtable}{0.5\textwidth}
		%		\centering
		%		\begin{tabular}{l c c c c}
		%			\midrule
		%			\textbf{Size} & \textbf{Image} & \textbf{Width} & \textbf{Layers} & \textbf{Neurons}\\
		%			\hline\\[-1.75ex]
		%			Small & 32 & 3 & 6 & 16 \\
		%			Medium & 48 & 3 & 6 & 32 \\
		%			Large & 64 & 3 & 6 &  48 \\
		%			\midrule
		%		\end{tabular}
		%		\caption{PathNet}
		%		\label{tbl:pathnet-params}
		%	\end{subtable}

		\begin{subtable}{0.3\textwidth}
			\centering
			\scalebox{0.85}{
				\begin{tabular}{l c c}
					\midrule
					\textbf{Size} & \textbf{Sequence} & \textbf{Neurons} \\
					\hline\\[-1.75ex]
					Small & 20 & 128 \\
					Medium & 30 & 512 \\
					Large & 40 & 1024 \\
					\midrule
				\end{tabular}
			}
			\caption{LSTM/PhasedLSTM}
			\label{tbl:lstm-params}	
		\end{subtable}%
		\begin{subtable}{0.3\textwidth}
			\centering
			\scalebox{0.85}{
				\begin{tabular}{l c c}
					\midrule
					\textbf{Size} & \textbf{Image} & \textbf{Neurons}\\
					\hline\\[-1.75ex]
					Small & 32 & 16 \\
					Medium & 48 & 32 \\
					Large & 64 & 48 \\
					\midrule
				\end{tabular}
			}
			\caption{PathNet}
			\label{tbl:pathnet-params}
		\end{subtable}
		\begin{subtable}{0.3\textwidth}
			\centering
			\scalebox{0.85}{
				\begin{tabular}{l c c}
					\midrule
					\textbf{Size} & \textbf{Image} & \textbf{Width}\\
					\hline\\[-1.75ex]
					Small & 128 & 1 \\
					Medium & 192 & 2 \\
					Large & 256 & 4 \\
					\midrule
				\end{tabular}
			}
			\caption{GoogleNet}
			\label{tbl:googlenet-params}
		\end{subtable}
	\end{adjustwidth}
	\caption{Parameters of the deep learning models in evaluation.}
	\label{tbl:model-params}
	\vspace{-1ex}
\end{table*}

\noindent\textbf{Environment}
In the evaluation, we ran \graphi with various experiment settings and compared its performance with TensorFlow (version $1.2.0$), a popular state-of-the-art deep learning framework on Intel Xeon Phi processor 7250 (see Section~\ref{sec:background}). \graphi compiled operations of the computation graph via ICC (version $17.0.4 20170411$), and links to Intel MKL 2017 for matrix multiplication and LIBXSMM (version $1.8.1$) for convolution. TensorFlow is configured to run multiple operations in parallel on the manycore CPU using Intel MKL 2017 for matrix multiplication and convolution. 

%However, the execution engine has not been carefully tailored for the manycore CPU.

%Titan X Pascal has 3584 CUDA cores with base frequency 1417 MHz, and in total achieves around 11 TFLOPS peak performance. The machine equipped Titan X Pascal also has two Intel Xeon E5-2683 v3 @ 2.00GHz processors (previously code named Haswell), each with 14 cores, and they communicate with the GPU via PCI-E 3.0.

\noindent\textbf{Deep learning models} The experiments ran the training phases of four neural networks. The first one is LSTM~\cite{hochreiter1997long}, a popular recurrent neural network model with applications in modeling text~\cite{sutskever2014sequence, sutskever2011generating}, speech~\cite{graves2013speech, fan2014tts}, and video~\cite{yue2015beyond, patraucean2015spatio}. 
%These data can all be viewed as sequence data, and LSTM maintains ``memory'' states based on previous steps to help make predictions about the next step, which is able to capture long term dependencies in the data.
Both our implementation and the TensorFLow benchmark of LSTM are based on~\cite{zaremba2014recurrent}.

The second is PhasedLSTM~\cite{neil2016phased}, a recent variant of LSTM which suits for processing asynchronous sensory events that carry timing information.
%It adds a gate that updates each memory state at a trainable periodic cycle, and is well suited for processing asynchronous sensory events that carry timing information.
TensorFlow provides PhasedLSTM cell for benchmarking, and we implemented an identical counterpart for \graphi. Note that the customized optimizations for LSTM cannot be easily applied to PhasedLSTM even if these two networks only have slight difference. However, the optimizations with \graphi apply to both since it is neural network agnostic.

%TensorFlow also provides a monolithic implementation of LSTM cell, which does slightly better for smaller networks. However, for the same reason as above we use the basic version for evaluation, and similar optimizations, e.g. fusing multiple operations to reduce overhead, is orthogonal to our efforts and can be applied on \graphi as well. 

Table~\ref{tbl:lstm-params} summarizes the three network sizes of LTSM-like networks following the Tensorflow convention.

%TensorFlow benchmark suggests three sizes of 4-layer LSTM neural network by varying the total length of sequence, and the number of neurons in each layer. We adopted this setting for our experiments with LSTM and PhasedLSTM in the evaluation, with is summarized in Table~\ref{tbl:lstm-params}.

The third one we used is a convolutional neural network (CNN) called PathNet~\cite{fernando2017pathnet} invented by DeepMind. PathNet is designed to be trained on multiple tasks simultaneously, leading to many parallel modules in each layer.
%PathNet is inspired by the empirical observation that one module can be shared by multiple tasks, and the whole network is able to learn multiple tasks faster leveraging these common building blocks. 
We implemented this neural network for both \graphi and TensorFlow using one $3\times 3$ convolution, followed by rectified linear units and a $2\times 2$ pooling in each module.
%The key parameters of the overall model are the input image size, the number of layers, the active modules in one layer for each task, and the output neurons of the convolution operation in each module. 
We chose 3 sizes for evaluation based on the original study, with number of layers set to 3, active modules per layer set to 6, and the remaining parameters summarized in Table~\ref{tbl:pathnet-params}.

Lastly, we also evaluated on GoogleNet~\cite{szegedy2015going}, a deep CNN model widely used in image classification. It does not have as many parallel operations compared to the previous 3 networks, so there is less room for optimization in \graphi, nevertheless, the ``inception'' modules in GoogleNet still consists of 2-3 parallel convolution/pooling operations, so sequential execution is suboptimal. We refer to the implementation provided in TensorFlow for evaluation as well, but vary the image size and multiply the number of output filters in each convolution by a constant factor (i.e. ``width'' of the network~\cite{ZagoruykoK16}) to obtain models of 3 different sizes (Table~\ref{tbl:googlenet-params}).

For LSTM/PhasedLSTM/PathNet, the batch size is set to 64, and for GoogleNet, the batch size is 32, to maximally utilize the 16GB MCDRAM.

\begin{figure*}[htbp]
	\centering
	\begin{subfigure}{0.45\textwidth}
		\centering
		\includegraphics[width=\textwidth]{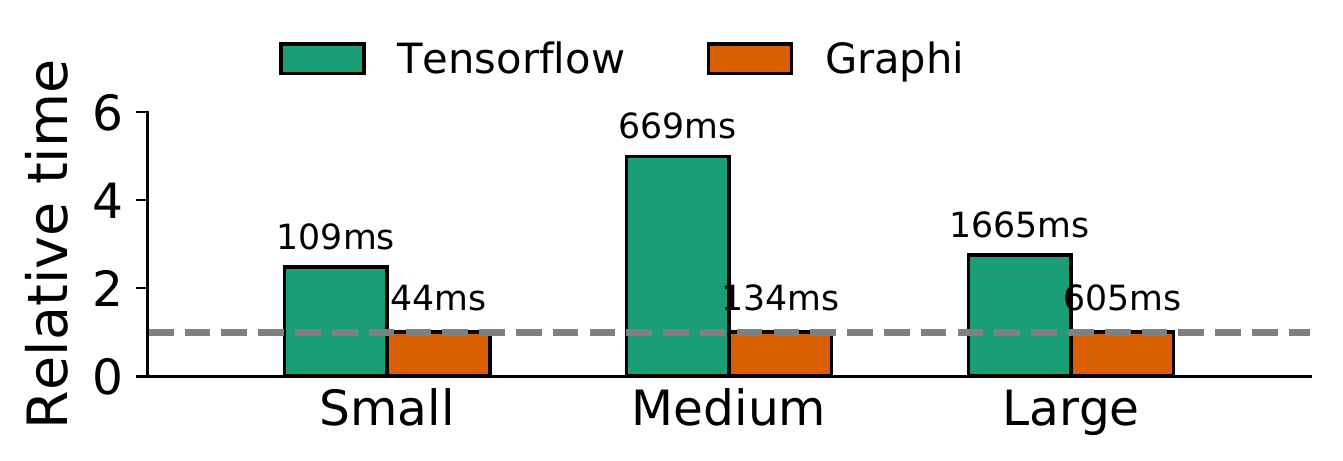}
		\caption{LSTM}
		\label{fig:lstm-training}
	\end{subfigure}
	~
	\begin{subfigure}{0.45\textwidth}
		\centering
		\includegraphics[width=\textwidth]{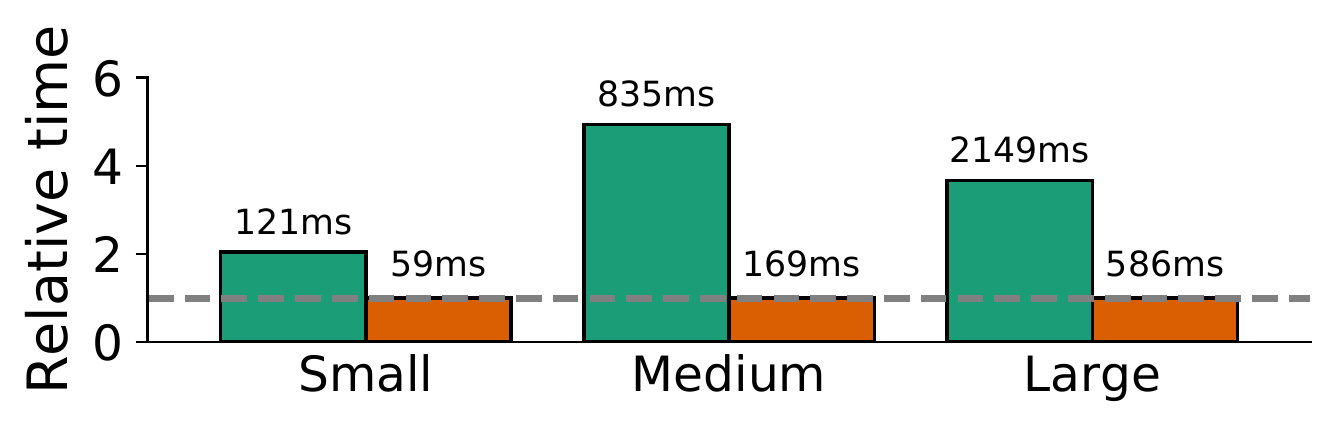}
		\caption{PhasedLSTM}
		\label{fig:plstm-training}
	\end{subfigure}
	\\
	\begin{subfigure}{0.45\textwidth}
		\centering
		\includegraphics[width=\textwidth]{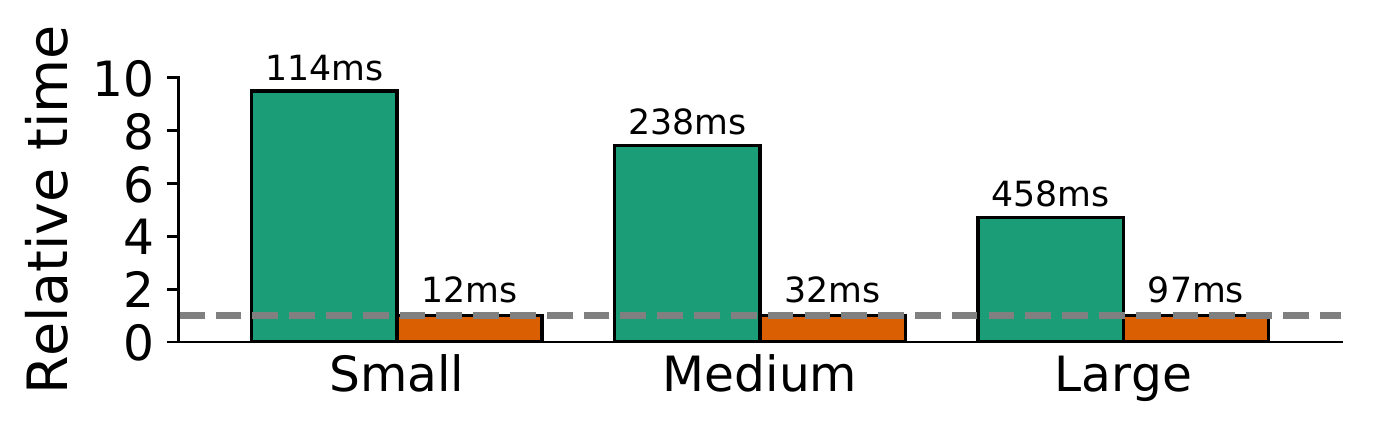}
		\caption{PathNet}
		\label{fig:pathnet-training}
	\end{subfigure}
	~
	\begin{subfigure}{0.45\textwidth}
		\centering
		\includegraphics[width=\textwidth]{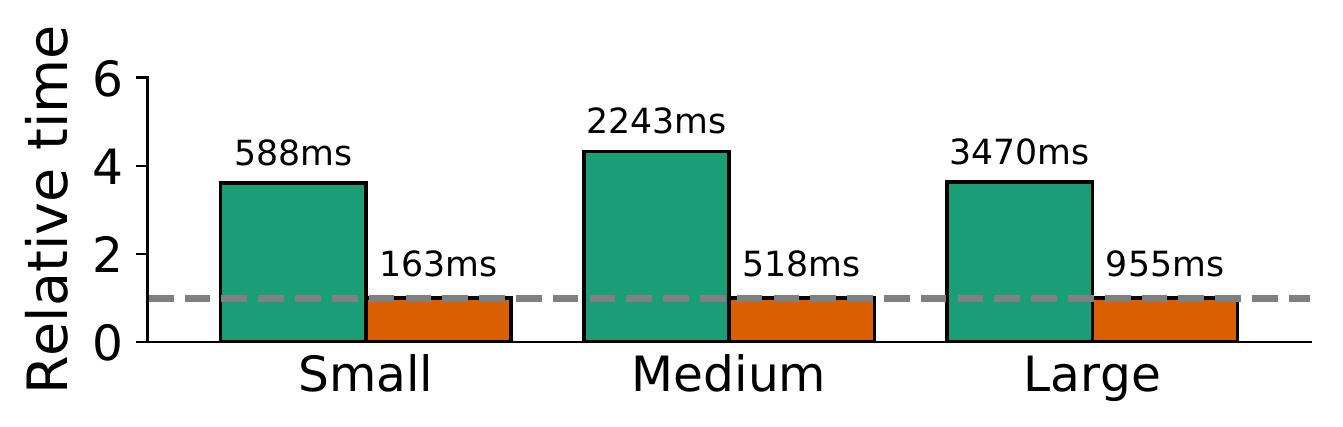}
		\caption{GoogleNet}
		\label{fig:googlenet-training}
	\end{subfigure}
	\caption{Batch training time of TensorFlow and \graphi on the manycore CPU. y-axis shows the relative running time to \graphi, lower is better.}
	\vspace{-1em}
	\label{fig:overall-results}
\end{figure*}

\subsection{Overall Results}
\label{sec:eval:overall}

Figure~\ref{fig:overall-results} shows the overall results of batch training times of different models by both \graphi and TensorFlow.  These are results of the best parallelization settings for both \graphi and TensorFlow.   For clarity, we normalized the batch training time of different models.

The results show that \graphi achieves 2.1-9.5$\times$ speed-up compared with TensorFlow on the manycore CPU. For LSTM/PhasedLSTM, because both \graphi and TensorFlow relies on MKL for the time consuming matrix multiplication operations, the better results are largely attributed to \graphi's execution engine. Since TensorFlow does not control thread placement, multiple threads often run on the same physical core, causing interference and unpredictable performance. Besides, TensorFlow uses Eigen for element-wise operations, which has its own thread pool, making the problem worse. 

Moreover, Eigen divides all the element-wise operations into small chunks and manages them in a centralized job queue. This causes contention as well, and we think this helps to explain why \graphi performed best with the medium sized networks relative to the small/large: for the small networks, each operation is not divided into many chunks and the effect is not too damaging; and for the large networks, Eigen's job queue design is less of a bottleneck because each operation now takes longer to run.

For PathNet, \graphi achieved about 9.5$\times$ speed-up on the large size, 7$\times$ on the medium, and 4$\times$ on the small. Aside from the difference of execution engines, we think this is also attributed to the building primitives, because LIBXSMM, the library \graphi uses for convolution, has been specially optimized for small convolutions compared with the Intel MKL convolution implementation.

\graphi was about 3-4$\times$ faster on GoogleNet of all three different sizes. Although GoogleNet is a relatively simple network with less optimization opportunities, \graphi still benefits from the better parallel scheduling, in addition to the more performant primitives provided by LIBXSMM.

In the next two subsections, we try to decompose the
speedup contribution made by our proposed schemes. We
analyzed the effect of parallel execution (\ref{sec:eval:parallel}) and intelligent scheduling (\ref{sec:eval:scheduler}), and the rest is attributed to the elimination of resource interference.

\begin{comment}
Although Intel Xeon Phi processor 7250 has fewer peak FLOPS than \nvidia Titan X Pascal GPU, \graphi did quite well on the small networks in comparison with TensorFlow running on GPU. It is 30\% lower with LSTM, but 2$\times$ faster with PhasedLSTM, and has similar performance with PathNet. This is because TensorFlow runs sequentially on GPU, a single small operation cannot effectively utilize  many GPU cores. However, \graphi's advantage generally diminishes when the network size increases, as the bigger operations can better utilize the whole GPU, and the manycore CPU has less computing power.

Note that although LSTM and PhasedLSTM are conceptually similar, their specific implementations are different, with a larger number of smaller and parallelizable operations in PhasedLSTM, favoring the design of \graphi. In addition, we find that Intel MKL does not perform well for the matrix multiplication operations in LSTM compared with cuBLAS on the GPU, attributing much to the low performance of \graphi on the large-sized LSTM. In addition, we notice that when running PhasedLSTM on the GPU machine, the host processors are also used up to 30\%. It turns out that the \emph{modulo} operator presented in the design of PhasedLSTM (to compute the periodical cycles) is not supported on the GPU. So the data has to be transmitted to CPUs for computation, which hurts the overall performance. Although the manycore CPU's design is not mature yet, we think this highlights some of its potential advantage of being more general-purpose than GPU.
\end{comment}

\subsection{Varying number of executors}
\label{sec:eval:parallel}

\begin{figure*}[htbp]
	\centering
	\begin{subfigure}{0.45\textwidth}
		\centering
		\includegraphics[width=\textwidth]{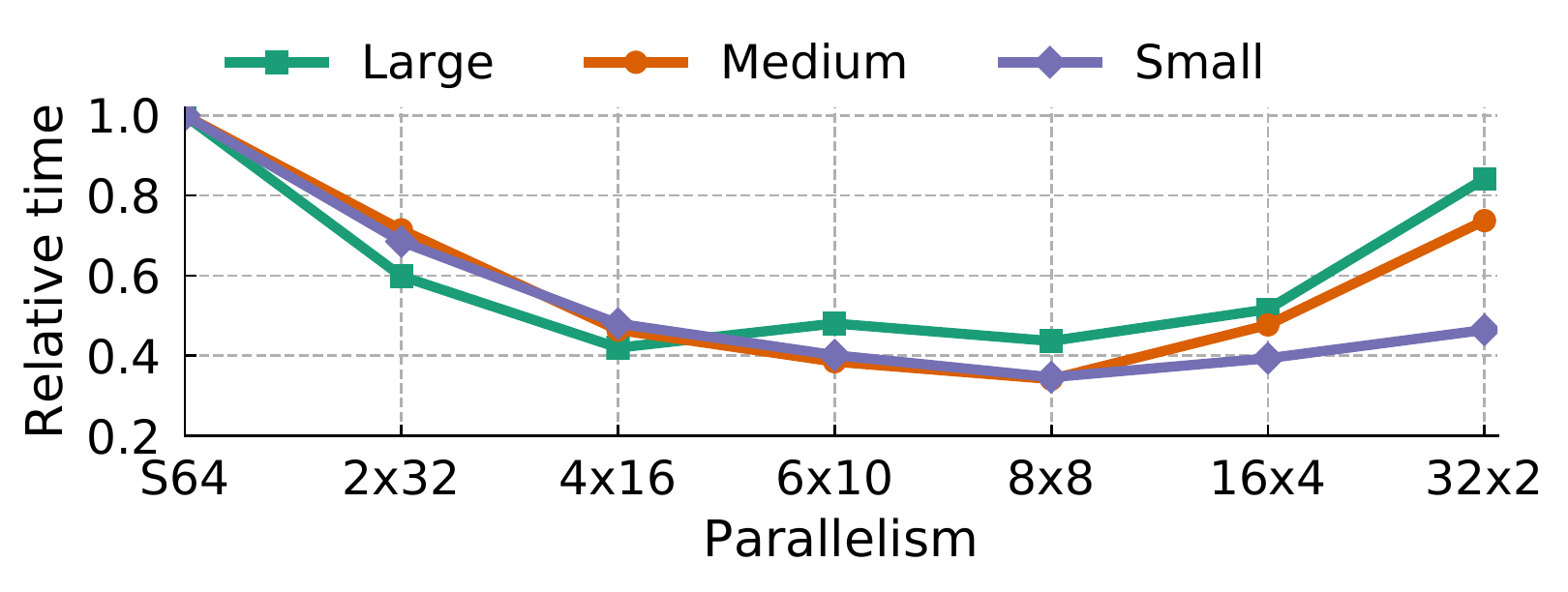}
		\caption{LSTM}
		\label{fig:lstm-parallelism}
	\end{subfigure}
	~
	\begin{subfigure}{0.45\textwidth}
		\centering
		\includegraphics[width=\textwidth]{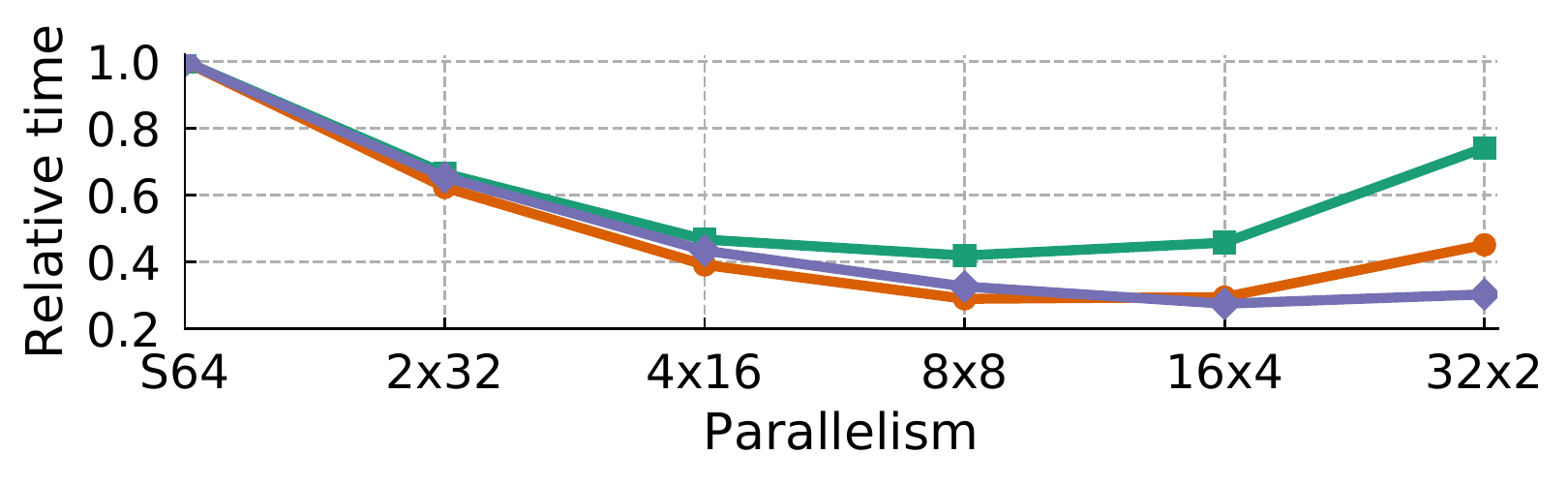}
		\caption{PhasedLSTM}
		\label{fig:plstm-parallelism}
	\end{subfigure}
	\\
	\begin{subfigure}{0.45\textwidth}
		\centering
		\includegraphics[width=\textwidth]{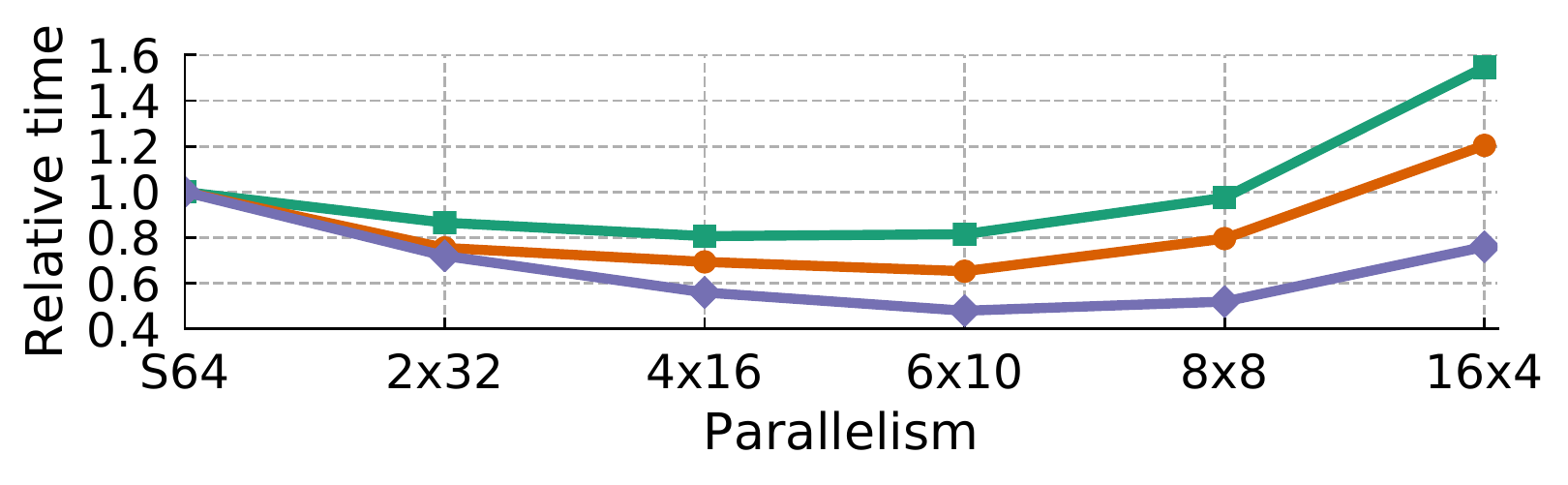}
		\caption{PathNet}
		\label{fig:pathnet-parallelism}
	\end{subfigure}
	~
	\begin{subfigure}{0.45\textwidth}
		\centering
		\includegraphics[width=\textwidth]{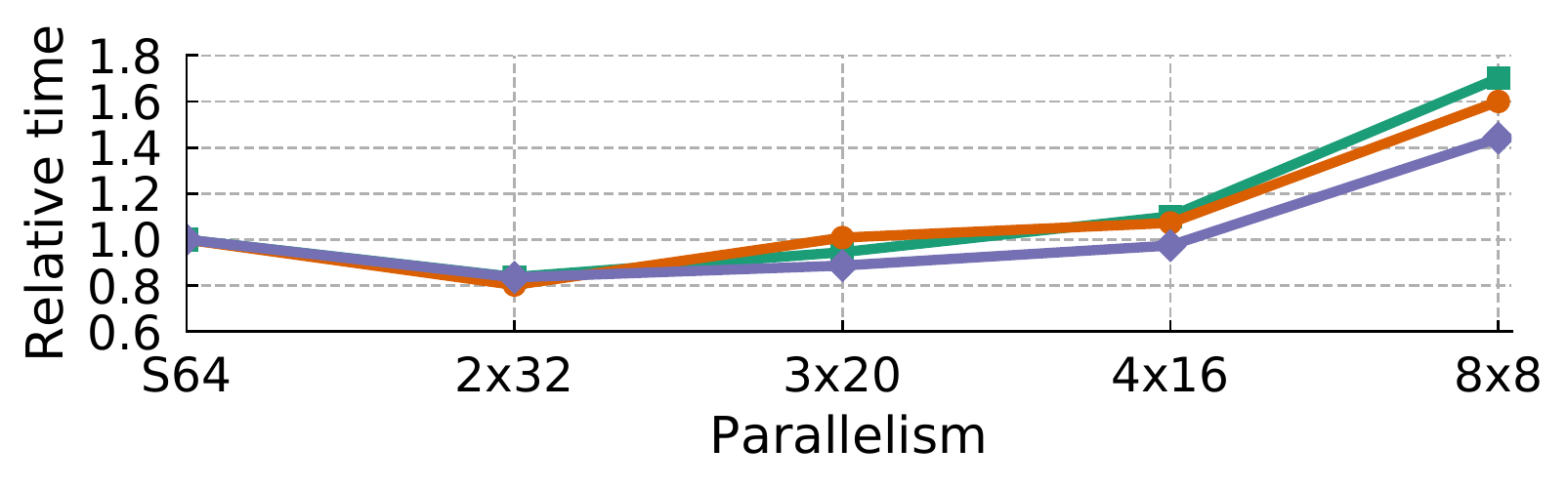}
		\caption{GoogleNet}
		\label{fig:googlenet-parallelism}
	\end{subfigure}
	\caption{Relative batch training time of \graphi under different parallelism settings compared with a sequential execution engine. $\bm{n\times k}$ on x-axis means $\bm{n}$ executors, each using $\bm{k}$ cores. S64 refers to sequential execution engine using 64 cores.}
	\label{fig:parallelism}
    %\vspace{-3em}
\end{figure*}

In this experiment we varied the number of executors in \graphi when running different neural networks, and observed how the performance changed. The Intel Xeon Phi processor 7250 has $68$ cores, $2$ of which was reserved for the scheduler and the light-weight executor, respectively as discussed in Section~\ref{sec:impl}.   We varied the number of executors from $k=2, 4, 8, 16, 32$, and assigned $64/k$ cores to each executor. PathNet in our setting had 6 modules per layer, so in addition to the aforementioned settings, we added one setting with $6$ executors, each using $10$ cores. And for GoogleNet because it has $2$-$3$ parallel operations, we also tried $3$ executors each with $10$ cores. 
%We omit results beyond $16\times4$/$8\times 8$ for these networks, respectively, because there is not enough parallelism and performance has declined significantly before then. 
For comparison, we also ran a sequential execution engine.
%which sorted the operations topologically beforehand and ran them sequentially with a simple loop in each iteration.
Figure~\ref{fig:parallelism} shows the relative batch training times of \graphi compared with the sequential engine.

From the figure we can see that parallel executions of \graphi achieved significant speed-ups for all 4 models on Intel Xeon Phi processor 7250. For LSTM and PhasedLSTM, the highest improvements ranged from 2.3$\times$ to 3.1$\times$, and for PathNet, they ranged from 1.2$\times$ to 2.1$\times$. In general, the speed-ups for small networks are more pronounced, because the small operations in these networks lead to poor usage of the cores with a sequential execution engine. For GoogleNet with less parallelism opportunities, the speed up is smaller and around 1.2$\times$, and the performance decreases rapidly when we have more than 2 executors.

Figure~\ref{fig:parallelism} also shows that different numbers of executors were needed to achieve the maximum performance for different networks. This highlights the importance of the profiling step (Section~\ref{sec:design:profiler}) of \graphi since the optimal number of parallel executors is related to the structure of the model. Specifically, in the four-layer LSTM/PhasedLSTM model, one cell from each layer can run in parallel, and there are 2-3 parallel operators in each cell, so the total number of parallelizable operations is around 8-12. PathNet/GoogleNet model has 6/2-3 modules in each layer that can run in parallel, respectively, corresponding exactly to the number of optimal number of executors needed.

When surpassing the optimal setting, the performance starts to decrease, with the large networks suffering most, because there are not enough parallel operations to utilize all the executors simultaneously, resulting in some executors being idle most of the time.

Note that although we have enumerated through the configurations to obtain optimal number of executors needed in these experiments, in practice, it is also possible to infer some good settings through static analysis from the graph structure, just like what we have done above.
%it can also be done automatically by enumerating the settings in the initial iterations. %Because the same graph typically run for many thousands of iterations in model training/evaluation, the subsequent iterations and can enjoy the speed up.

\subsection{\graphi scheduler}

\begin{table}[tbp]
	\centering
	\resizebox{0.5\textwidth}{!}{%
		\begin{tabular}{l c c c c}
			\midrule
			\textbf{Parallelism} & \textbf{LSTM} & \textbf{PhasedLSTM} & \textbf{PathNet} & \textbf{GoogleNet} \\
			\hline\\[-1.75ex]
			2$\times$32 & 0.86 & 0.81 & 0.88 & 0.94 \\
			4$\times$16 & 0.88 & 0.85 &  0.92 & 0.96 \\
			8$\times$8 & 0.82 & 0.91 & 0.89 & 0.93 \\
			16$\times$4 & 0.91 & 0.86 & 0.91 & 0.91 \\
			32$\times$2 & 0.87 & 0.85 & 0.92 & 0.92 \\
			\midrule
		\end{tabular}
	}
	\caption{Relative batch training time of \graphi vs. naive parallel scheduler on medium-sized networks.}
	\label{tbl:graphi-speed-up}
	\vspace{-2em}
\end{table}

\label{sec:eval:scheduler}
Table~\ref{tbl:graphi-speed-up} summarizes the relative training time of \graphi compared with the naive scheduling used in TensorFlow and MXNet on medium-sized networks under various parallelism configurations. Note that in this comparison, we have eliminated all executor thread interference, so the performance difference only comes from the scheduler. For the sake of space, we only show the speed-up on medium sized networks, and the results on small/large networks are consistent.

\graphi achieved 8\%-19\% speed-up compared with the naive scheduling, in which all executors independently poll the centralized queue for operations. When the number of executors is large (e.g. in the manycore CPU), the heavy concurrent use of the centralized queue causes contention. In addition, part of improvement is also attributed to the critical-path first scheduling based on the operation-level profiling.

%In addition, the simple scheduler does not have global knowledge of the graph.  The scheduler has to schedule arbitrary operations to arbitrary executors in a greedy manner. In contrast, \graphi's centralized scheduler is able to take the global computation graph information into consideration and come up with more advanced scheduling scheme (in our case, critical-path-first scheduling). 

Specifically, there was greater speed-up on LSTM and PhasedLSTM because they have many more small operations, which results in severer contention on the global queue in the naive scheduling. Correspondingly, the improvement on GoogleNet is smaller because each operation is larger, resulting in less contention when polling the queue. Moreover, the more complex structures of the LSTM/PhasedLSTM computation graphs, the greater gain critical-path-first scheduling provides. In effect, the hand-optimized LSTM implementation by cuDNN~\cite{cudnn-lstm} follows a diagonal parallel execution pattern for the LSTM cells on different layers and sequence locations. We visualized the operation execution trace and found the critical-path scheduler recovered the same pattern automatically (details omitted due to space limit), while the baseline scheduler failed to.

\section{Related Works}
\label{sec:related}
Modern deep learning frameworks mostly express the computation of deep learning models in computation graphs. Caffe~\cite{jia2014caffe} and neon~\cite{neon} use computation graphs with layers and explicitly labels the forward and backward paths, which limits the overall optimization opportunities. TensorFlow~\cite{abadi2016tensorflow}, MXNet~\cite{chen2015mxnet}, Theano~\cite{bergstra2010theano} and Caffe2~\cite{caffe2} all express the computation in the pure computation graph. Such a design method leaves more room for optimization. 

By default, the execution engines of the existing deep learning frameworks execute the computation graph in sequence according to its topological order. This is conceptually simple and feasible to neural networks with large operation since the within-operation parallelization is able to fully utilize the computation resources. For neural networks with smaller operations and more complex structures, TensorFlow and MXNet provide parallel execution engines for CPUs. However, their simple scheduling and thread interference often result in sub-optimal performance, especially on the manycore CPU. The \graphi execution engine proposed in this paper surmounts these two limitations to obtain high performance for various neural networks on the hardware.

The idea of expressing the computation as a directed acyclic graph (or in some context, dependence graph) dates back to the data flow computation~\cite{Dennis:1974:PAB:642089.642111,dennis1980data} in 1970s. Much of the later work focused on optimizing the execution performance in different scenarios especially in the compilers~\cite{kuck1981dependence,ottenstein1984program,padua1986advanced,ferrante1987program,cytron1991efficiently}. The work in this paper leverages the previous idea and applies it to a new application (deep learning computation graph) on manycore CPU architecture. 

The online scheduling problem, i.e. how to dynamically schedule $M$ jobs with dependencies to run on $N$ workers has been studied for decades, including theoretical results such as achievable upper bounds of an online algorithm~\cite{feldmann1993optimal} as well as heuristic greedy solutions~\cite{hu1961parallel,coffman1972optimal}. The \graphi scheduler design is inspired mainly by the critical-path first scheduling algorithm~\cite{hu1961parallel} and the multiple queue skiplist scheduler~\cite{muqss}.

How to eliminate thread interference has been an important performance optimization issue on multicore or manycore CPUs. Much of previous work focused on mapping different applications to different cores~\cite{das2013application} or partitioning memory for multiple applications~\cite{muralidhara2011reducing,liu2012software}. Our work follows a similar rule-of-thumb to assure that different threads use disjoint resources so as to maximally reduce the resource contention in the \graphi execution engine.

The manycore CPU used in our experiments runs as an independent host.  However, a manycore CPU can be viewed as an accelerator. Related work on optimizations for deep learning computations on the accelerators include GPU~\cite{chetlur2014cudnn}, FPGA~\cite{farabet2009cnp,zhang2015optimizing,zhang2017frequency,han2017ese} and the first generation Intel Xeon Phi coprocessor (Knights Corner)~\cite{zlateski2016znn}.  Their focus was on the optimizations of the critical deep learning primitives, e.g. convolution and matrix multiplication, which are the operations of a computation graph. Our work, in contrast, aims at optimizing the entire computation graph, and focuses more on inter-operation optimizations. 

For specific neural networks like LSTM, customized optimizations such as operation fusion, network pruning and parallel execution of the computation graph (similar to \graphi) exist, including optimizations for GPUs~\cite{cudnn-lstm} and FPGAs~\cite{han2017ese}. However, these are ad-hoc solutions specific to certain kinds of neural networks. Classical neural network like LSTM has many variants such as time-frequency LSTM~\cite{sainath2016modeling}, grid LSTM~\cite{kalchbrenner2015grid}, LSTM with layer normalization~\cite{ba2016layer} and dropout~\cite{semeniuta2016recurrent}, phased LSTM~\cite{neil2016phased}, group LSTM~\cite{kuchaiev2017factorization}, etc. \graphi is a generic execution engine, optimized for all variations without specializations.  

\section{Conclusions}
\label{sec:concl}
This paper proposes \graphi, a generic and high-performance execution engine for running computation graphs of deep learning models on manycore CPUs.  The focus of our work is on efficiently executing deep learning computation graphs, especially ones with small operations and complex structures.  To achieve this goal, \graphi automatically finds the optimal parallel setting with profiling, minimizes the interference of parallel operations, and further improves execution efficiency with critical-path first scheduling. 

Our experiments with four different neural networks showed that \graphi outperformed TensorFlow (optimized for CPU via MKL) by 2.1$\times$ to 9.5$\times$ on Intel Xeon Phi processor 7250. Further detailed analysis shows that proper parallelism without interference improves performance by 1.2$\times$ to 3.1$\times$, and better scheduling results in 8\% to 19\% speed up and automatically recovers a handcrafted parallelization scheme for LSTM in cuDNN.

The work in this paper is a first step towards designing an efficient parallel engine for deep learning computation graph execution on the manycore CPU. Beyond the scope of this paper, we also have verified that \graphi achieves favarbale speedup on the latest multicore CPUs (Intel Xeon Platinum 8180, code named Skylake), demonstrating the generalizability of our framework. There are several future directions, including applying the ideas to GPUs and FPGAs as well as the distributed systems with multiple nodes, extending \graphi to handle dynamic computation graphs, and further optimizing \graphi for challenging memory hierarchies such as NUMA.

%\pagebreak

\paragraph{Acknowledgements}
The authors thank Alex Heinecke, Pradeep Dubey, Hans Pabst, Niranjan Hasabnis from Intel Corporation, Zhen Jia from Princeton University, Jie Gao from Stony Brook University, Yungang Bao from Institute of Computing Technology Chinese Academy of Sciences, and Xiaoqiang Zheng from Google for valuable discussions.
%\newpage

\bibliographystyle{ACM-Reference-Format}
\bibliography{references}

\end{document}